\documentclass[prd,aps,twocolumn,10pt,letterpaper,superscriptaddress]{revtex4-2}
\usepackage[utf8]{inputenc}
\usepackage[english]{babel}
\usepackage{bm}
\usepackage{graphicx}
\usepackage{amsmath}
\usepackage{amssymb}
\usepackage{amsfonts}
\usepackage{xcolor}
\usepackage{braket}
\usepackage{hyperref}
\usepackage{multirow}
\usepackage{float}
\hypersetup{
    pdfstartview={FitH},    % fits the width of the page to the window
    colorlinks=true,       % false: boxed links; true: colored links
    linkcolor=blue,          % color of internal links
    citecolor=blue,        % color of links to bibliography
    filecolor=blue,      % color of file links
    urlcolor=blue           % color of external links
}

\begin{document}

%\title{Three-dimensional angle-resolved distributions of positrons created in supercritical collisions of heavy nuclei considering rotational coupling}
\title{Three-dimensional calculations of positron creation in supercritical collisions of heavy nuclei}

\author{N. K. Dulaev}
\email{st069071@student.spbu.ru}
\affiliation{Department of Physics,
St.~Petersburg State University, 7-9 Universitetskaya nab., St.~Petersburg 
199034, Russia}

\author{D. A. Telnov}
\email{d.telnov@spbu.ru}
\affiliation{Department of Physics,
St.~Petersburg State University, 7-9 Universitetskaya nab., St.~Petersburg 
199034, Russia}

\author{V. M. Shabaev}
\email{v.shabaev@spbu.ru}
\affiliation{Department of Physics,
St.~Petersburg State University, 7-9 Universitetskaya nab., St.~Petersburg 
199034, Russia}
\affiliation{Petersburg Nuclear Physics Institute named by B.~P.~Konstantinov of National Research Center ''Kurchatov Institute'', Orlova roscha 1, 188300 Gatchina, Leningrad region$,$ Russia}

\author{Y.~S.~Kozhedub}
\affiliation{Department of Physics,
St.~Petersburg State University, 7-9 Universitetskaya nab., St.~Petersburg 
199034, Russia}

\author{X.~Ma}
\affiliation{Institute of Modern Physics,
Chinese Academy of Sciences, 730000 Lanzhou, China}

\author{I. A. Maltsev}
\affiliation{Department of Physics,
St.~Petersburg State University, 7-9 Universitetskaya nab., St.~Petersburg 
199034, Russia}

\author{R. V. Popov}
\affiliation{Petersburg Nuclear Physics Institute named by B.~P.~Konstantinov of National Research Center ''Kurchatov Institute'', Orlova roscha 1, 188300 Gatchina, Leningrad region$,$ Russia}
\affiliation{Department of Physics,
St.~Petersburg State University, 7-9 Universitetskaya nab., St.~Petersburg 
199034, Russia}

\author{I. I. Tupitsyn}
\affiliation{Department of Physics,
St.~Petersburg State University, 7-9 Universitetskaya nab., St.~Petersburg 
199034, Russia}

\begin{abstract}
Energy--angle differential and total probabilities of positron creation in slow supercritical collisions of two identical heavy nuclei are calculated beyond the monopole approximation. The time-dependent Dirac equation (TDDE) for positrons is solved using the generalized pseudospectral method in modified prolate spheroidal coordinates, which are well-suited for description of close collisions in two-center quantum systems. In the frame of reference where the quasimolecular axis is fixed, the rotational coupling term is added to the Hamiltonian. Unlike our previous calculations, we do not discard this term and  retain it when solving the TDDE. Both three-dimensional angle-resolved and angle-integrated energy distributions of outgoing positrons are obtained. Three-dimensional angle-resolved distributions exhibit a high degree of isotropy. For the collision energies in the interval 6 to 8 MeV/u, the influence of the rotational coupling on the distributions and total positron creation probabilities is quite small. 
\end{abstract}

\maketitle

\section{Introduction}
Slow collisions of heavy nuclei provide a valuable tool for exploring the nonperturbative regime of quantum electrodynamics (QED) in strong electromagnetic field. In Ref. \cite{Pomeranchuk_1945}, where properties of hydrogenlike ions depending on nuclear charge number $Z$ were studied, it was demonstrated that at a certain critical $Z_{\mathrm{cr}}$ the $1s$ state reaches the negative-energy electron continuum. Subsequent studies by Soviet and German physicists \cite{Gershtein_1969, Pieper_1969_Interior, Popov_1970_1, *Popov_1970_2, *Popov_1970_3, *Popov_1971_1, Zeldovich_1971_Electronic, Muller_1972_Solution, *Muller_1972_Electron, Mur_1976, Popov_1976, Muller_1976_Positron, Reinhardt_1977_Quantum, Soff_1977_Shakeoff, Rafelski_1978_Fermions, Greiner_1985_Quantum} have shown that for $Z > 173$ the initially unoccupied $1s$ state dives into the negative continuum as a resonance, decaying with some probability with emission of one or two electron-positron pairs. Since it is currently impossible to produce such heavy nuclei, heavy ion collisions with the charge numbers $Z_1 + Z_2 > Z_{\mathrm{cr}}$ are considered instead. When sufficiently heavy nuclei approach each other within a critical distance $R_{\mathrm{cr}}$, the quasimolecular level $1s\sigma_g$ dives into the electron negative-energy continuum for a short period (on the order of $10^{-21}$ s), leading to spontaneous vacuum decay into charged vacuum and positrons, which can escape the nuclei and be detected.  This process, known as spontaneous vacuum decay in heavy-ion collisions, is discussed in more detail in, e.g., Refs. \cite{Greiner_1985_Quantum, Rafelski_2017_Probing} and references therein. 
A major challenge in the experimental and theoretical study of this phenomenon is that the spontaneous pair-creation mechanism is strongly masked by dynamic process arising from the time-dependent variation of the electromagnetic potential of the nuclei.

Numerous theoretical studies have been conducted on the subject over the years. In static approximation the calculations of pair production were carried out in Refs.~\cite{Popov_1973, Peitz_1973_Autoionization, Popov_1979}. The supercritical resonances were examined in Refs.~\cite {Ackad_2007_Numerical, Godunov_2017_Resonances, Ackad_2007_Supercritical, Marsman_2011_Calculation, Maltsev_2020_Calculation, Zaytsev_2014_QED}. The dynamics of the pair production in collisions of heavy nuclei was extensively studied by the Frankfurt group (see, e.g.,~\cite{Smith_1974_Induced, Reinhardt_1981_Theory, Muller_1988_Positron, Rafelski_1978_Fermions, Greiner_1985_Quantum, Bosch_1986_Positron, Muller_1994_Electron, Reinhardt_2005_Supercritical, Rafelski_2017_Probing}). In the monopole approximation spontaneous and dynamic channels of the pair production have been evaluated in Refs. \cite{Maltsev_2015_Electron, Bondarev_2015_Positron, Maltsev_2019_How, Popov_2020_How, Reus_2022_Positron}, and beyond the monopole approximation in Refs. \cite{Maltsev_2017_Pair, Maltsev_2018_Electron, Popov_2018_One, Popov_2023_Spontaneous, Dulaev_2024_Angular}. The QED-vacuum polarization effects in the supercritical Coulomb field have been examined in Refs.~\cite{Grashin_2022_Vacuum, Krasnov_2022_Non}. In Ref. \cite{Voskresensky_2021_Electron}, the relativistic semiclassical approach was applied to study the instability of electron-positron vacuum.

Recently, our research group proposed a method to distinguish the subcritical pair-production process from the supercritical one \cite{Maltsev_2019_How, Popov_2020_How}. This approach uses a specific type of nuclear trajectories characterized by the minimal internuclear distance $R_{\mathrm{min}}$ and scaled energy parameter $\varepsilon \geq 1$,
\begin{equation}
 \varepsilon = E / E_0,
 \label{eq:scaled_E}
\end{equation}
where $E$ is the collision energy and $E_0$ is the head-on collision energy. Along trajectories with fixed $R_{\mathrm{min}}$ and various $\varepsilon$, the nuclei produce the same field strength. As $\varepsilon$ decreases, the duration of the supercritical regime increases, enhancing the contribution of spontaneous mechanism, while the nuclei velocity and the dynamical channel are reduced. Thus, an increase in pair-creation probability as $\varepsilon$ approaches 1 is a signature of the supercritical regime. This signature was observed in one-center calculations within \cite{Maltsev_2019_How, Popov_2020_How} and beyond \cite{Popov_2023_Spontaneous} the monopole approximation. The same behavior was confirmed in two-center calculations in Ref. \cite{Dulaev_2024_Angular}.

With new opportunities for experimental research on heavy ion collisions at upcoming facilities \cite{Gumberidze_2009_X, Akopian_2015_Layout, Lestinsky_2016_Physics, Ma_2017_HIAF}, it is important to investigate not only the total pair-creation probabilities but also angle-resolved energy distributions. In Ref. \cite{Dulaev_2024_Angular}, two-dimensional angular-energy positron distributions in the collision plane were analyzed, demonstrating that the distributions are nearly spherically symmetric with a high degree of isotropy. These calculations were performed without including the rotational term (see, e.g., \cite{Muller_1976_The}), as it was previously argued that the effect of the rotation coupling on the total pair creation should be small \cite{Betz_1976_Direct, Soff_1979_Electrons, Reinhardt_1980_Dynamical}. However, further investigation is needed to assess the impact of the rotational term on the angle-resolved probability distributions.

This work aims to study the impact of the rotational term on the total pair-creation probabilities, as well as on the angle-integrated and angle-resolved positron energy distributions, in slow collisions of heavy nuclei. The time-dependent Dirac equation is solved using the generalized pseudospectral method in modified prolate spheroidal coordinates, with the rotational term included in the Hamiltonian. The magnetic field of the nuclei is neglected, as its influence is expected to be negligible  \cite{Soff_1988_Positron}. To obtain angle-resolved energy distributions the plane wave decomposition approach is employed. The three-dimensional angle-resolved energy distributions of positrons are analyzed.  

Atomic units ($\hbar=|e|=m_{e}=1$) are used throughout the paper unless specified otherwise.

\section{Methods}\label{methods}
Theoretical methods which we use to treat the problem were described in detail in our previous paper~\cite{Dulaev_2024_Angular}. Here, we recall the basic expressions with the emphasis on the rotational coupling term in the Hamiltonian, which was discarded in the previous calculations~\cite{Popov_2023_Spontaneous, Dulaev_2024_Angular}. Making use of Dirac's hole picture~\cite{Greiner_1985_Quantum,Godunov_2017_Resonances}, we start with the time-dependent Dirac equation (TDDE) for a positron in the field of two colliding bare nuclei in the center-of-mass frame of reference:
\begin{equation}
    i \frac{\partial}{\partial t} \Psi(\bm{r}, t) = H\Psi(\bm{r},t), \label{eq:tdde}
\end{equation}
where $\Psi(\bm{r},t)$ is a four-component wave function of the positron, and the Hamiltonian $H$ is written as
\begin{equation}
\begin{split}
    H &= c(\bm{\alpha} \cdot \bm{p}) + c^2 \beta\\ 
    &+ U_{n}(|\bm{r}-\bm{a}(t)|) + U_{n}(|\bm{r}+\bm{a}(t)|).
\end{split}
    \label{eq:hamiltonian}
\end{equation}
In Eq.~(\ref{eq:hamiltonian}),  $c$ is the speed of light, $\bm{p}$ is the momentum operator, $\bm{\alpha}$ and $\beta$ are the Dirac matrices. The function $U_{n}(r)$ represents a spherically symmetric nuclear potential within the extended nucleus model. We adopt the Fermi model of the nuclear charge distribution \cite{Elton_1961_Nuclear}, and the root-mean-square radii of the nuclei are taken from Ref. \cite{Angeli_2013_Table}. The vectors $\pm\bm{a}(t)$ define the instantaneous positions of the nuclei during the collision, so $2a(t)=R(t)$ where $R(t)$ is the internuclear distance. In the course of collision, both the absolute value and direction of the vector $\bm{a}$ vary with time. We assume that the internuclear axis of the quasimolecule is initially directed along the axis, which we denote as $z_{0}$ axis,
and the collision takes place in the $z_{0}-x_{0}$ plane. During the collision the internuclear axis is rotated about the $y$ axis which is perpendicular to the collision plane. Upon the unitary transformation of the wave function,
\begin{equation}
    \Psi(\bm{r}, t) = \exp{\left[ -i\chi J_y \right]} \Psi^{(r)}(\bm{r},t),\label{eq:wv_in_rot}
\end{equation}
where $\chi$ is the rotation angle of the internuclear axis ($\chi\rightarrow 0$ as $t\rightarrow -\infty$) and $J_y$ is the total angular momentum projection operator onto the $y$ axis, equal to the sum of the orbital and spin angular momentum projection operators $L_y$ and $S_y$:
\begin{equation}
    J_y = L_y + S_y,
\label{eq:ang_mom_proj}
\end{equation}
we arrive at the equation for the wave function $\Psi^{(r)}(\bm{r},t)$:
\begin{equation}
\begin{split}
    &i \frac{\partial \Psi^{(r)}(\bm{r}, t)}{\partial t} = \Big[c(\bm{\alpha} \cdot \bm{p}) + c^2 \beta - \dot{\chi} J_y\\
    &+U_{n}(|\bm{r}-a(t)\bm{e}_{z}|) + U_{n}(|\bm{r}+a(t)\bm{e}_{z}|)
\Big] \Psi^{(r)}(\bm{r}, t).
\end{split}
\label{eq:eq_rotated}
\end{equation}
where $\bm{e}_{z}$ is a unit vector along the internuclear axis, which is rotated by the angle $\chi$ from $z_{0}$ axis. We will term the transformation (\ref{eq:wv_in_rot}) a transition to the rotating frame of reference and $\Psi^{(r)}(\bm{r},t)$ the wave function in the rotating frame of reference, respectively, although Eq.~(\ref{eq:eq_rotated}) differs slightly from the equation in the relativistic rotating frame of reference (for details, see  Ref.~\cite{Muller_1976_The}). 

In the rotating frame of reference, the nuclei are located on the $z$ axis at all times, but the rotational coupling term $-\dot{\chi} J_y$ is introduced in the TDDE. This operator couples the states if their angular momentum projections onto the internuclear ($z$) axis differ by $\pm 1$. In our previous works~\cite{Popov_2023_Spontaneous, Dulaev_2024_Angular} the rotational coupling term was discarded, since it was argued in Refs.~\cite{Betz_1976_Direct, Soff_1979_Electrons, Reinhardt_1980_Dynamical} that the influence of this term on the total production of positrons in slow collisions of heavy nuclei is negligible. In the present study, we retain the rotational coupling term in Eq.~(\ref{eq:eq_rotated}) and assess its influence on the total positron creation probabilities as well as distributions of outgoing positrons with respect to their energies and emission angles.

It is convenient to perform an additional unitary transformation of the wave function in Eq.~(\ref{eq:eq_rotated}):
\begin{equation}
    \Psi^{(r)}=\begin{pmatrix}
        1&0&0&0\\
        0&\exp(i\varphi)&0&0 \\
        0&0&i&0 \\
        0&0&0&i\exp(i\varphi)
    \end{pmatrix}\psi ,
    \label{eq:phi_transform}
\end{equation}
where the angle $\varphi$ describes rotation about the $z$ axis. Upon this transformation, the positron state with the angular momentum projection $m+\frac{1}{2}$ onto the $z$ axis is described by the wave function $\psi^{(m)}\exp(im\varphi)$ where all four components of the function $\psi^{(m)}$ depend on the time as well as cylindrical coordinates $\rho$ and $z$ but not on $\varphi$. In the general case, the wave function $\psi(\bm{r},t)$ is represented by the series expansion
\begin{equation}
 \psi(\bm{r},t) = \sum\limits_{m=-\infty}^{\infty} \psi^{(m)}(\rho,z,t)\exp(im\varphi),
 \label{eq:m_expansion}
\end{equation}
and the time-dependent equation for the function $\psi(\bm{r},t)$ is recast in a set of coupled equations for the angular momentum projection components  $\psi^{(m)}$:
\begin{equation}
    i \frac{\partial}{\partial t} \psi^{(m)} = \mathsf{H}^{(m)}\psi^{(m)} - \sum\limits_{m'=m\pm 1} \dot{\chi}\mathsf{J}_y^{(m,m')} \psi^{(m')}. \label{eq:tdde_aft_trans}
\end{equation}
Except the rotational coupling term, the Dirac Hamiltonian in Eq.~(\ref{eq:eq_rotated}) is invariant under rotation about the $z$ axis and  diagonal in the angular momentum projection representation. Thus the Hamiltonians $\mathsf{H}^{(m)}$ in Eq.~(\ref{eq:tdde_aft_trans}) include the kinetic and rest energy as well as interaction with the nuclei for the states with the angular momentum projection $m+\frac{1}{2}$. The states with different angular momentum projections are coupled by the rotational coupling operator $\mathsf{J}_y$, which is modified from the conventional definition in Eq.~(\ref{eq:ang_mom_proj}) due to the transformation (\ref{eq:phi_transform}):
\begin{equation}
    \mathsf{J}_y = \mathsf{L}_y + \mathsf{S}_y, \label{eq:jy}
\end{equation}

\begin{equation}
    \mathsf{L}_y = L_y - \frac{z}{\rho} \sin{\varphi} 
    \begin{pmatrix}
        0 & 0 & 0 & 0 \\
        0 & 1 & 0 & 0 \\
        0 & 0 & 0 & 0 \\
        0 & 0 & 0 & 1 
    \end{pmatrix}, \label{eq:ly}
\end{equation}
\begin{equation}
  L_y =-i\cos\varphi\left(z\frac{\partial}{\partial\rho}-\rho\frac{\partial}{\partial z}\right) +i\frac{z}{\rho}\sin\varphi\frac{\partial}{\partial\varphi},
  \label{eq:l_y}
\end{equation}
\begin{equation}
\begin{split}
    \mathsf{S}_y &= \frac{i}{2} \exp(-i\varphi) 
    \begin{pmatrix}
        0 & 0 & 0 & 0 \\
        1 & 0 & 0 & 0 \\
        0 & 0 & 0 & 0 \\
        0 & 0 & 1 & 0 
    \end{pmatrix}\\
&-\frac{i}{2} \exp(i\varphi) 
    \begin{pmatrix}
        0 & 1 & 0 & 0 \\
        0 & 0 & 0 & 0 \\
        0 & 0 & 0 & 1 \\
        0 & 0 & 0 & 0 
    \end{pmatrix}    .\label{eq:sy}
\end{split}
\end{equation}

We use cylindrical coordinates in Eqs.~(\ref{eq:ly})--(\ref{eq:sy}). Two-center quantum systems, however, are naturally described with prolate spheroidal coordinates; for close collisions in such systems, prolate spheroidal coordinates can be slightly modified to make a smooth transition to a one-center system in the united-atom limit. Our modified prolate spheroidal coordinates $\lambda$ and $\eta$ are related to the cylindrical coordinates $\rho$ and $z$ as follows:
\begin{equation}
\begin{split}
    \rho  &= \sqrt{ \left[ (\lambda + a)^2 - a^2 \right] \left[ 1 - \eta^2 \right] }, \\
    z  &= (\lambda + a) \eta\\
    &(0\le\lambda<\infty, -1\le\eta\le 1). \label{eq:mpsc}
\end{split}
\end{equation}
The azimuthal angle $\varphi$ has the same definition in both coordinate systems. In the limit $a\rightarrow 0$, the coordinate $\lambda$ becomes the spherical radial coordinate $r$, and the coordinate $\eta$ is equal to $\cos\vartheta$ where $\vartheta$ is the polar angle in the spherical coordinate system with the axis $z$ directed along the internuclear axis.

Since the coordinate transformation (\ref{eq:mpsc}) is time-dependent through the parameter $a(t)$, the set of equations (\ref{eq:tdde_aft_trans}) is further modified:
\begin{equation}
\begin{split}
    i \frac{\partial}{\partial t} \psi^{(m)}(\lambda,\eta,t) = \left[\mathsf{H}^{(m)}-\frac{\dot{a}}{a}S_{a}\right]\psi^{(m)}(\lambda,\eta,t)\\ - \sum\limits_{m'=m\pm 1} \dot{\chi}\mathsf{J}_y^{(m,m')} \psi^{(m')}(\lambda,\eta,t), 
\end{split}
    \label{eq:tdde_final}
\end{equation}
with the scaling operator $S_{a}$ defined as 
\begin{equation}
 S_{a} = ia\frac{\partial\lambda}{\partial a}\frac{\partial}{\partial \lambda} + ia\frac{\partial\eta}{\partial a}\frac{\partial}{\partial \eta}.
\label{eq:scaling}
\end{equation}
In Eq.~(\ref{eq:scaling}), the partial derivatives of $\lambda$ and $\eta$ with respect to $a$ are calculated from the equations (\ref{eq:mpsc}) at fixed $\rho$ and $z$.

To solve the final set of the time-dependent equations (\ref{eq:tdde_final}), we employ the generalized pseudospectral (GPS) method, proved to be accurate and efficient for both nonrelativistic and relativistic problems~\cite{Telnov_2007_Ab, Telnov_2009_Effects, Telnov_2018_Multiphoton, Telnov_2021_Hyd}. Upon the GPS discretization, the operators on the right-hand side of (\ref{eq:tdde_final}) are transformed to a block-structure matrix. The diagonal blocks are formed by the $\mathsf{H}^{(m)}$ and $S_{a}$ matrices while the off-diagonal blocks are due to the rotational coupling $\mathsf{J}_y^{(m,m')}$. The linear dimension of each block is equal to $4\times N_{\lambda}\times N_{\eta}$ with $N_{\lambda}$ and $N_{\eta}$ being the numbers of grid points for the coordinates $\lambda$ and $\eta$, respectively. In the present calculations, we use $N_{\lambda}=320$, $N_{\eta}=20$ and retain the angular momentum projections $m$ from $-1$ to $1$. Thus the linear dimension of the total matrix in the set (\ref{eq:tdde_final}) is equal to $76800$. The box size for the pseudoradial coordinate $\lambda$ is chosen as $60/Z$ a.u. where $Z$ is the nuclear charge number. For the time propagation, we adopt the Crank--Nicolson algorithm \cite{Crank_1947_A} with a non-uniform time step where the smallest time step corresponds to the closest approach of the nuclei. The total number of time steps is equal to $2048$. 

To obtain energy--angle distributions of outgoing positrons, the unbound positron wave packet at the end of the time propagation is projected onto the relativistic plane waves. The relativistic plane wave functions $\Psi_{\bm{k}}^{(1)}$ and $\Psi_{\bm{k}}^{(2)}$ with the momentum $\bm{k}$, corresponding to the spin projection $\frac{1}{2}$ and $-\frac{1}{2}$, respectively, onto the $z$ axis in the rest frame of the positron can be written as \cite{bere1982}:
\begin{equation}
    \Psi_{\bm{k}}^{(1)} = 
    \begin{pmatrix}
    \sqrt{E_{k} + 2c^2} \\
    0 \\
    \sqrt{E_{k}} \cos{\vartheta_k} \\
    \sqrt{E_{k}} \sin{\vartheta_k} \, \exp(i \varphi_k)
    \end{pmatrix}
    \frac{\exp{[i (\bm{k}\cdot\bm{r}) - iE_{k}t]}}{(2\pi)^{3/2}\sqrt{2E_{k} + 2c^2}},
    \label{eq:pw1}
\end{equation}
\begin{equation}
    \Psi_{\bm{k}}^{(2)} = 
    \begin{pmatrix}
    0 \\
    \sqrt{E_{k} + 2c^2} \, \exp(i \varphi_k)\\
    \sqrt{E_{k}}  \sin{\vartheta_k} \\
    -\sqrt{E_{k}} \cos{\vartheta_k}\, \exp(i \varphi_k)
    \end{pmatrix}
    \frac{\exp{[i (\bm{k}\cdot\bm{r}) - iE_{k}t]}}{(2\pi)^{3/2}\sqrt{2E_{k} + 2c^2}}.
    \label{eq:pw2}
\end{equation}
In Eqs.~(\ref{eq:pw1}) and (\ref{eq:pw2}), $\vartheta_k$ and $\varphi_k$ are the angles which define the direction of the momentum vector $\bm{k}$ in the spherical coordinate system, and $E_{k}$ is the positron kinetic energy related to the momentum $k$ as follows:
\begin{equation}
    k = \frac{1}{c} \sqrt{E_{k}(E_{k} + 2c^2)}.
\end{equation}
The differential probabilities of positron creation in the states $\Psi_{\bm{k}}^{(1)}$ and $\Psi_{\bm{k}}^{(2)}$ are given by the following equations:
\begin{equation}
    \frac{dP^{(1)}}{dE_{k}d\Omega}=\frac{1}{c^3} \sqrt{E_{k}(E_{k}+2c^2)} (E_{k} + c^2) |\braket{\Psi_{\bm{k}}^{(1)}|\Psi^{(c)}}|^2, \label{eq:p1}
\end{equation}
\begin{equation}
    \frac{dP^{(2)}}{dE_{k}d\Omega}=\frac{1}{c^3} \sqrt{E_{k}(E_{k}+2c^2)} (E_{k} + c^2) |\braket{\Psi_{\bm{k}}^{(2)}|\Psi^{(c)}}|^2, \label{eq:p2}
\end{equation}
where $\Psi^{(c)}$ is the unbound wave packet which is calculated by projecting the positron wave function at the end of the time propagation onto the positive-energy continuum. 
In the frames of reference other than the rest frame of the positron, the spin projection is not a good quantum number, and the positron creation differential probability without the spin selection is obtained as a sum of the differential probabilities (\ref{eq:p1}) and (\ref{eq:p2}):
\begin{equation}
  \frac{dP}{dE_{k}d\Omega} =  \frac{dP^{(1)}}{dE_{k}d\Omega} +  \frac{dP^{(2)}}{dE_{k}d\Omega}.   \label{eq:distr}
\end{equation}

\section{Results}\label{results}
We performed calculations of positron creation in slow collisions of two identical bare nuclei with the charge numbers $Z=92$ and $Z=96$. Collisions with the scaled energies $\varepsilon=1.0,$ $1.1,$ $1.2,$ $1.3$ were studied. Since the velocities of the nuclei are nonrelativistic, their motion follows classical Rutherford trajectories. The time-dependent Dirac equation is solved as described in Sec. \ref{methods} for $6$ initial discrete vacuum positron states with the angular momentum projection $J_z=1/2$, lying most closely to the onset of the positive-energy continuum. Transitions from these states to the positive-energy continuum make a dominant contribution to the positron creation. The initial states with $J_z=-1/2$ contribute equally, thus only the initial states with $J_z=1/2$ are propagated, and the resulting positron creation probability is multiplied by $2$. 

To study the effect of the rotational coupling, two sets of calculations were performed. In the first set, the rotational coupling term $-\dot{\chi} \mathsf{J}_y$ was discarded, thus the angular momentum projection $J_{z}=1/2$ was conserved in the \textit{rotating frame of reference}. In the second set, we included the states with  $J_{z}=-1/2$, $1/2$ and $3/2$. When the nuclei approach each other during the collision, the adiabatic potential curves of the quasimolecules may exhibit avoided crossing leading to nonadiabatic transitions between the adiabatic states. To account for this effect, the initial and final internuclear separations were set at $R_{\mathrm{max}} = 5.5/Z$ a.u. since avoided crossings between the most important for the positron creation adiabatic quasimolecular states $1s_{1/2}$, $2p_{1/2}$, $2s_{1/2}$, $3p_{1/2}$, and $3s_{1/2}$ (notation refers to the united atom limit) occur at smaller internuclear distances. As in our previous works, we choose the minimum internuclear separation $R_{\mathrm{min}}$ equal to $17.5$~fm. At this internuclear distance, the highest-lying discrete positron states $1s_{1/2}$ for $Z=92$ and $Z=96$ are already supercritical resonances deeply in the upper positron continuum. 

\subsection{Total and angular-momentum-resolved positron creation probabilities}
Table \ref{table:total} presents the total positron creation probabilities for collisions with the values of $Z$ and $\varepsilon$ used in the present  calculations. The results without the rotational coupling term ($J_{z}=1/2$) and with the rotational coupling term and three angular momentum projections retained in the calculations ($J_{z'}=-1/2,$ $1/2$, $3/2$) are compared. The total probabilities are obtained by summing the contributions from the quasimolecular states with the respective angular momentum projections $J_{z}$. The table also lists the absolute differences in probabilities between the two sets of the calculations. As shown in the table, the relative difference in the probabilities inserted by the $-\dot{\chi} \mathsf{J}_y$ term is about $10^{-3}$. For collisions of the nuclei with $Z=92$, the relative difference does not exceed 0.1\%, while for $Z=96$ it is about 0.05\%. For head-on collisions with $\varepsilon=1.0$, transitions between the states with different angular momentum projections do not occur, only the states with $J_{z}=1/2$ are populated, so the calculations with and without the rotational coupling term give identical results.
\begin{table}
\centering
\setlength\tabcolsep{6pt}
\caption{\label{table:total} Total positron creation probabilities in collisions of two identical nuclei with $R_{\mathrm{min}}=17.5$ fm. The initial angular momentum projection of the positron states is $J_z=1/2$. Results of the calculations without the rotational coupling term ($J_{z}=1/2$) and with the rotational coupling term ($J_{z}=-1/2,$ $1/2$, $3/2$) are given in the third and fourth column, respectively. The last column shows the difference between the two calculations.}
\begin{ruledtabular}
\begin{tabular}{ccccc} 
\multirow{2}{*}{$Z$}  & \multirow{2}{*}{$\varepsilon$} & \multicolumn{2}{c}{$J_{z}$}   & \multirow{2}{*}{Difference} \\ \cline{3-4}
                    &                      & $1/2$       & $-1/2$, $1/2$, $3/2$ &                             \\ \hline
\multirow{4}{*}{92} & 1.0                  & $1.132 \times 10^{-2}$ & $1.132 \times 10^{-2}$      & $0$                    \\
                    & 1.1                  & $1.120 \times 10^{-2}$ & $1.121 \times 10^{-2}$      & $9.00 \times 10^{-6}$                    \\
                    & 1.2                  & $1.109 \times 10^{-2}$ & $1.109 \times 10^{-2}$      & $8.96 \times 10^{-6}$                    \\
                    & 1.3                  & $1.095 \times 10^{-2}$ & $1.096 \times 10^{-2}$      & $9.66 \times 10^{-6}$                    \\ \hline
\multirow{4}{*}{96} & 1.0                  & $4.111 \times 10^{-2}$ & $4.111 \times 10^{-2}$      & $0$                    \\
                    & 1.1                  & $3.895 \times 10^{-2}$ & $3.897 \times 10^{-2}$      & $1.32 \times 10^{-5}$                    \\
                    & 1.2                  & $3.729 \times 10^{-2}$ & $3.730 \times 10^{-2}$      & $1.52 \times 10^{-5}$                    \\
                    & 1.3                  & $3.581 \times 10^{-2}$ & $3.582 \times 10^{-2}$      & $1.74 \times 10^{-5}$      \\         
\end{tabular}
\end{ruledtabular}
\end{table}

As one can see, the effect of rotational coupling is in a slight increase of  the total positron creation probability for all $\varepsilon > 1.0$ used in the calculations. A possible explanation for this effect is that an increase of the number of channels for possible transitions generally increases the total transition probability, that is depletion of the initial state.
This effect is particularly noticeable in collisions with $Z=96$, where higher collision energies (compared to the case of $Z=92$) lead to a greater contribution from states with the momentum projections different from the initial projection. This is in agreement with increased rotation velocity of the internuclear axis. For collisions with $Z=92$, the effect of the rotational coupling is smaller, and the calculation inaccuracies make it difficult to clearly isolate an increase of this effect with the collision energy. However, the difference of the total positron creation probabilities due to the rotational coupling for $\varepsilon = 1.3$ is larger than for $\varepsilon = 1.1$, suggesting a slight increase of the rotational coupling influence at higher collision energies.

Table \ref{table:total} also suggests that accounting for the rotational coupling does not alter the previously found patterns in the dependence of the total probability on the scaled collision energy. In the supercritical regime, for collisions with $Z=92$ and $Z=96$, the total probability decreases with increasing $\varepsilon$, as expected. Such a behavior of the total positron creation probability proves a significant contribution of the spontaneous pair creation mechanism at $\varepsilon \rightarrow 1$.

As an additional verification of the present results obtained within two-center method in modified prolate spheroidal coordinates, we also performed calculations within the one-center approach beyond the monopole approximation, as outlined in Ref.~\cite{Popov_2023_Spontaneous}, with the rotational coupling term included. The calculations were carried out for U$^{92+}$--U$^{92+}$ collisions at $R_{\mathrm{min}}=17.5$~fm. These calculations confirm our results described above in that the rotational coupling has no significant effect on the total positron creation probabilities.

\begin{table}
\centering
\setlength\tabcolsep{3.5pt}
\caption{\label{table:projections} Contributions of different angular momentum projections $J_{z}$ to the total positron creation probabilities in collisions of two identical nuclei with $R_{\mathrm{min}}=17.5$ fm. The initial angular momentum projection of the positron states is $J_z=1/2$. Here $\gamma$ is the ratio of the contributions of states with $J_{z} = -1/2$ and $J_{z} = 1/2$ and $\nu$ is the kinematically predicted ratio.}
\begin{ruledtabular}
\begin{tabular}{ccccccc}
\multirow{2}{*}{$Z$}  & \multirow{2}{*}{$\varepsilon$} & \multicolumn{3}{c}{$J_{z}$} & \multirow{2}{*}{$\gamma$}              & \multirow{2}{*}{$\nu$} \\ \cline{3-5}
                    &                      & $-1/2$     & $1/2$      & $3/2$      & &                            \\ \hline
\multirow{4}{*}{92} & 1.0                  & $0$ & $1.13 \times 10^{-2}$ & $0$ & 0.00  & 0.00                       \\
                    & 1.1                  & $1.71 \times 10^{-3}$ & $3.90 \times 10^{-3}$ & $1.63 \times 10^{-6}$ & 0.44  & 0.44                       \\
                    & 1.2                  & $2.70 \times 10^{-3}$ & $2.84 \times 10^{-3}$ & $1.20 \times 10^{-6}$ & 0.95  & 0.96                       \\
                    & 1.3                  & $3.32 \times 10^{-3}$ & $2.15 \times 10^{-3}$ & $1.16 \times 10^{-6}$ & 1.54  & 1.56                       \\ \hline
\multirow{4}{*}{96} & 1.0                  & $0$ & $4.11 \times 10^{-2}$ & $0$ & 0.00  & 0.00                       \\
                    & 1.1                  & $5.92 \times 10^{-3}$ & $1.36 \times 10^{-2}$ & $1.98 \times 10^{-6}$ & 0.44  & 0.44                       \\
                    & 1.2                  & $9.09 \times 10^{-3}$ & $9.56 \times 10^{-3}$ & $1.50 \times 10^{-6}$ & 0.95  & 0.96                       \\
                    & 1.3                  & $1.09 \times 10^{-2}$ & $7.04 \times 10^{-3}$ & $1.45 \times 10^{-6}$ & 1.54  & 1.56  \\                 
\end{tabular}
\end{ruledtabular}
\end{table}
In Table \ref{table:projections}, we present the individual contributions of the states with different angular momentum projections to the total pair creation probability for the calculations with the rotational coupling term retained and angular momentum projections $J_{z}=-1/2$, $1/2$, and $3/2$ included. The column labeled $\gamma$ shows the ratio of the populations of the continuum positron states with $J_{z} = -1/2$ and $J_{z} = 1/2$. We note that these populations are obtained after the time propagation, when the internuclear ($z$) axis is in its final position rotated from the original direction by the angle $\chi_{\mathrm{max}}$. The value $\chi_{\mathrm{max}}$ can be obtained from solving the classical Rutherford scattering problem:
\begin{equation}
 \chi_{\mathrm{max}} = 2\arctan{\sqrt{4\varepsilon(\varepsilon-1)}}.
 \label{eq:chimax}
\end{equation}
As one can see, it depends only on the scaled collision energy and does not depend on the masses or charges of the colliding nuclei. Thus the same final rotation angles are obtained in collisions of the nuclei with $Z=92$ and $Z=96$. They are as follows:
\begin{equation}
\begin{split}
 \varepsilon &=1.1\quad \chi_{\mathrm{max}}=67.1^{\circ},\\
 \varepsilon &=1.2\quad \chi_{\mathrm{max}}=88.8^{\circ},\\
 \varepsilon &=1.3\quad \chi_{\mathrm{max}}=102.6^{\circ}.
\end{split}
\label{eq:chimax_val}
\end{equation}
We also note that the rotational coupling effect is different in the inertial and rotating frames of reference. Even if in the inertial frame the terms in the Hamiltonian which break the rotational symmetry about initial direction of the internuclear axis have negligible influence on the dynamics, so the angular momentum projection on this fixed axis is conserved, in the rotating frame the kinematic effect of the rotational coupling term is not small. This kinematic effect can be seen from the solution of a simple problem for the quantum-mechanical system with the angular momentum $\frac{1}{2}$. If such a system is in the state with the angular momentum projection $+\frac{1}{2}$ on the initial $z_{0}$ axis, then in the coordinate system rotated by the angle $\chi_{\mathrm{max}}$ about the $y$ axis, the probabilities to find the system in the states with the projections $+\frac{1}{2}$ and $-\frac{1}{2}$ on the rotated $z$ axis are equal to $\cos^{2}(\frac{1}{2}\chi_{\mathrm{max}})$ and $\sin^{2}(\frac{1}{2}\chi_{\mathrm{max}})$, respectively~\cite{Galitski_2013_Quantum}. Thus the kinematically predicted ratio of the populations of the continuum positron states with $J_{z} = -1/2$ and $J_{z} = 1/2$ must be equal to
\begin{equation} \label{eq:kin}
    \nu = \tan^2{\left(\frac{1}{2}\chi_{\mathrm{max}}\right)},
\end{equation}
if no transitions due to rotational coupling occur in the inertial frame.

The value of $\nu$ for different $\varepsilon$ is listed in the last column of Table \ref{table:projections}. As one can see for both $Z=92$ and $Z=96$, at each $\varepsilon$ the numerically calculated value $\gamma$ is close to the analytically predicted value of $\nu$. 

A slight difference between $\gamma$ and $\nu$ is observed at higher collision energies, $\varepsilon=1.2$ and $\varepsilon=1.3$. A possible reason of this difference may be in the dynamic effect of the rotational coupling at higher angular velocities of the internuclear axis, not accounted for in the kinematic formula (\ref{eq:kin}). Table \ref{table:projections} shows that states with $J_{z}=-1/2$ and $1/2$ contribute most significantly to the total probability, while the contributions of states with $J_{z}=3/2$ are $3-4$ orders of magnitude smaller. Since we did not perform calculations with larger numerical parameters, we cannot confirm that the numerical data for the $J_{z}=3/2$ contributions are reliable. However, the uncertainties of our calculations can be estimated as not exceeding those probabilities for the states with $J_{z}=3/2$.

\subsection{Angle-integrated energy distributions of outgoing positrons}
In the previous works \cite{Popov_2020_How, Popov_2023_Spontaneous, Dulaev_2024_Angular}, it was demonstrated within the monopole approximation and beyond that, in the supercritical regime, the maximum of the angle-integrated energy distributions of outgoing positrons decreases as $\varepsilon$ increases. In this section, we examine the positron energy spectra obtained in the two-center calculations with the rotational coupling included.  

\begin{figure}
    \centering
    \includegraphics[width=\columnwidth]{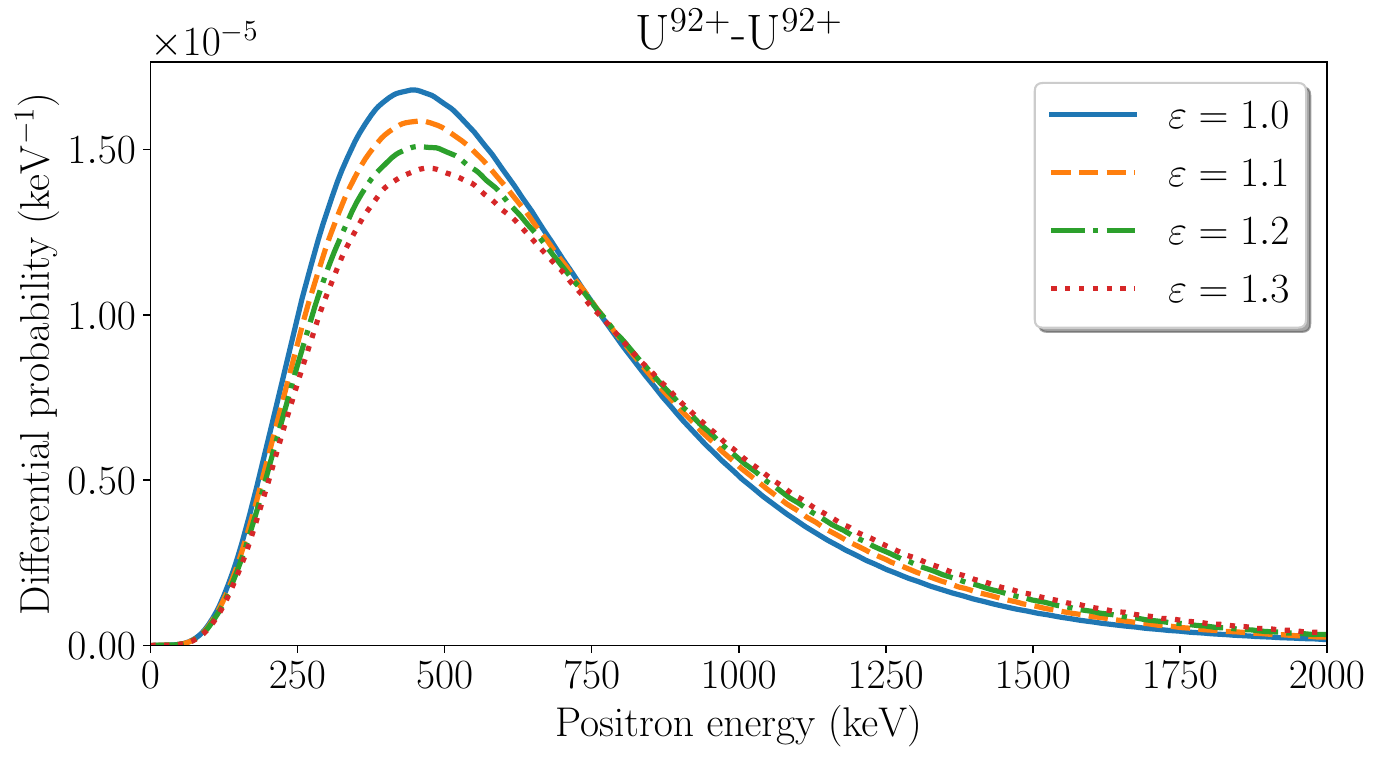}
    \caption{Energy spectra of positrons for symmetric collisions of nuclei with $Z=92$ at $R_{\mathrm{min}}=17.5$~fm and $\varepsilon=1.0$, $1.1$, $1.2$, $1.3$. Angular momentum projections $J_{z}=-1/2$, $1/2$, and $3/2$ are taken into account.}
    \label{fig:z_92}
\end{figure}

\begin{figure}
    \centering
    \includegraphics[width=\columnwidth]{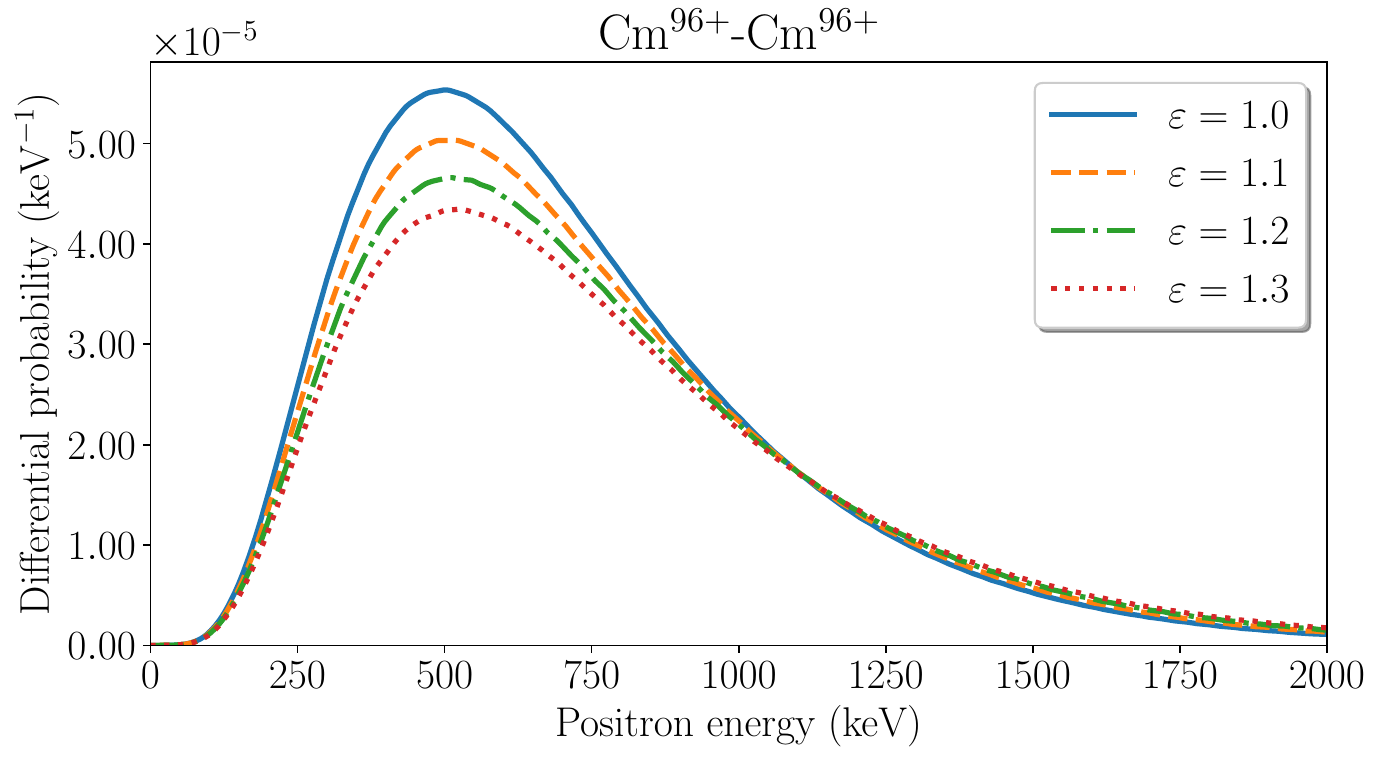}
    \caption{Energy spectra of positrons for symmetric collisions of nuclei with $Z=96$ at $R_{\mathrm{min}}=17.5$~fm and $\varepsilon=1.0$, $1.1$, $1.2$, $1.3$. Angular momentum projections $J_{z}=-1/2$, $1/2$, and $3/2$ are taken into account.}
    \label{fig:z_96}
\end{figure}

In Figs.~\ref{fig:z_92} and \ref{fig:z_96}, the positron energy spectra are shown for collisions of two identical nuclei with the charge numbers $Z=92$ and $Z=96$, respectively, and scaled energy parameters $\varepsilon=1.0$, $1.1$, $1.2$, $1.3$. Angular momentum projections $J_{z}=-1/2$, $1/2$ and $3/2$ were included in the time propagation of the initially selected $6$ discrete positron vacuum states with $J_{z}=1/2$, which make the most significant contributions to the positron creation probability. From the energy spectra in Figs.~\ref{fig:z_92} and \ref{fig:z_96} one can conclude that taking the rotational coupling into account when solving the TDDE does not alter qualitatively the pattern already seen in our previous calculations without the rotational coupling term~\cite{Popov_2020_How, Popov_2023_Spontaneous, Dulaev_2024_Angular}. In collisions with both $Z=92$ and $Z=96$, the maximum of the distribution is the highest at $\varepsilon=1.0$ and decreases as $\varepsilon$ increases to $1.3$. This observation is in a full agreement with the works~\cite{Popov_2020_How, Popov_2023_Spontaneous, Dulaev_2024_Angular} and is consistent with the conclusion made in the previous subsection that the influence of rotational coupling on the total positron creation probabilities is insignificant at the collision energies $6-8$~MeV/u studied in the present paper. We emphasize that this particular dependence of the energy distribution peak on the scaled collision energy signals about considerable contribution of the spontaneous mechanism to the electron-positron pair creation, as we discussed earlier~\cite{Popov_2020_How, Popov_2023_Spontaneous, Dulaev_2024_Angular}. Thus our present analysis of the rotational coupling effect confirms validity of our previously made observations about the signatures of the spontaneous positron creation in the energy distributions of outgoing positrons.

\subsection{Angle-resolved energy distributions of outgoing positrons}
\begin{figure}
    \centering
    \includegraphics[width=\columnwidth]{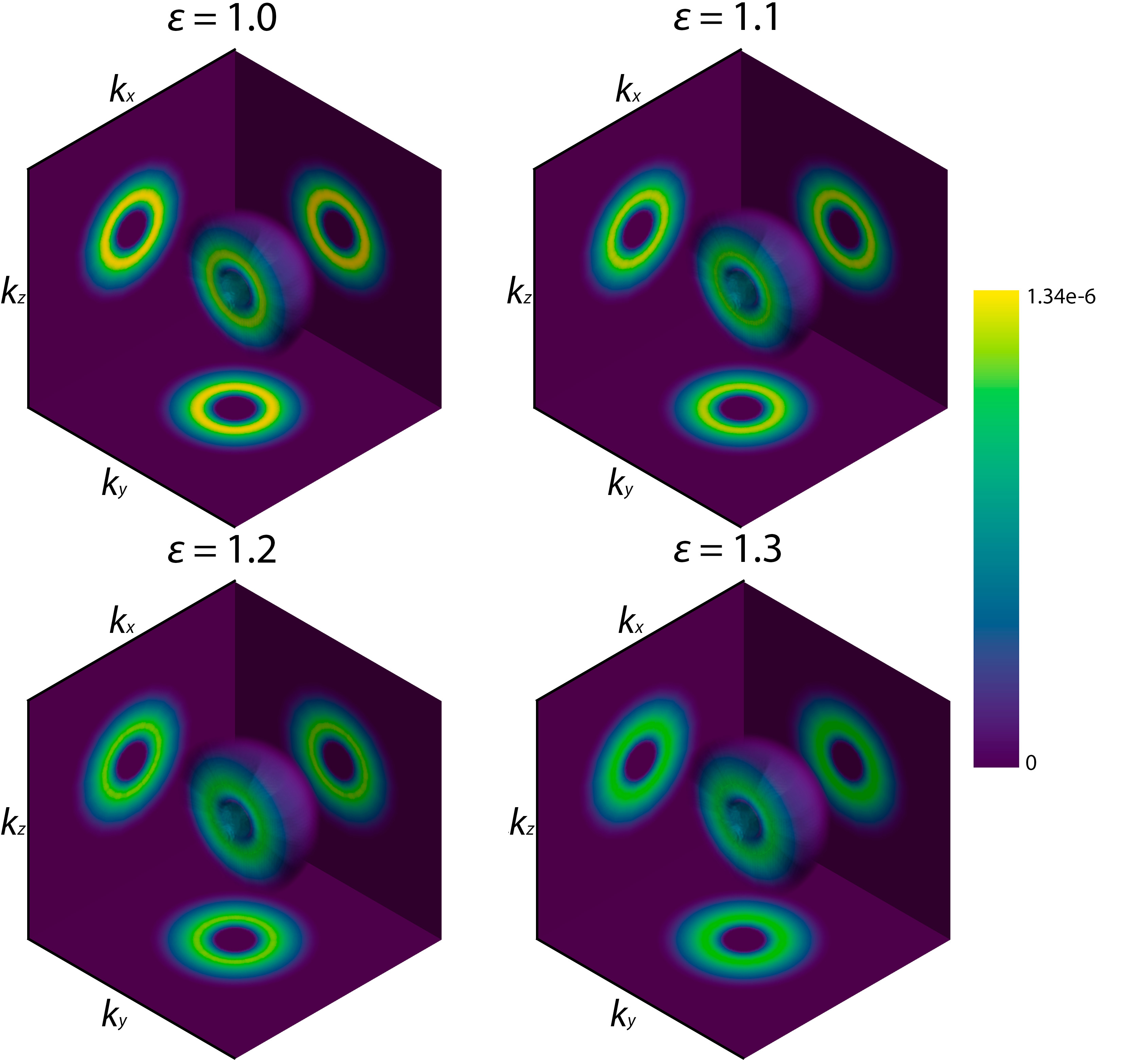}
    \caption{Pseudocolor plot of the angle-resolved differential probabilities of positron creation corresponding to the positron momentum components $k_z$, $k_x$, and $k_z$. The spectra are presented for collisions of the nuclei with $Z=92$ and $\varepsilon=1.0$, $1.1$, $1.2$, $1.3$ at $R_{\mathrm{min}}=17.5$~fm. Angular momentum projections $J_{z}=-1/2$, $1/2$ and $3/2$ are taken into account. For more details, please see the text.}
    \label{fig:3d_distributions_z_92}
\end{figure}

\begin{figure}
    \centering
    \includegraphics[width=\columnwidth]{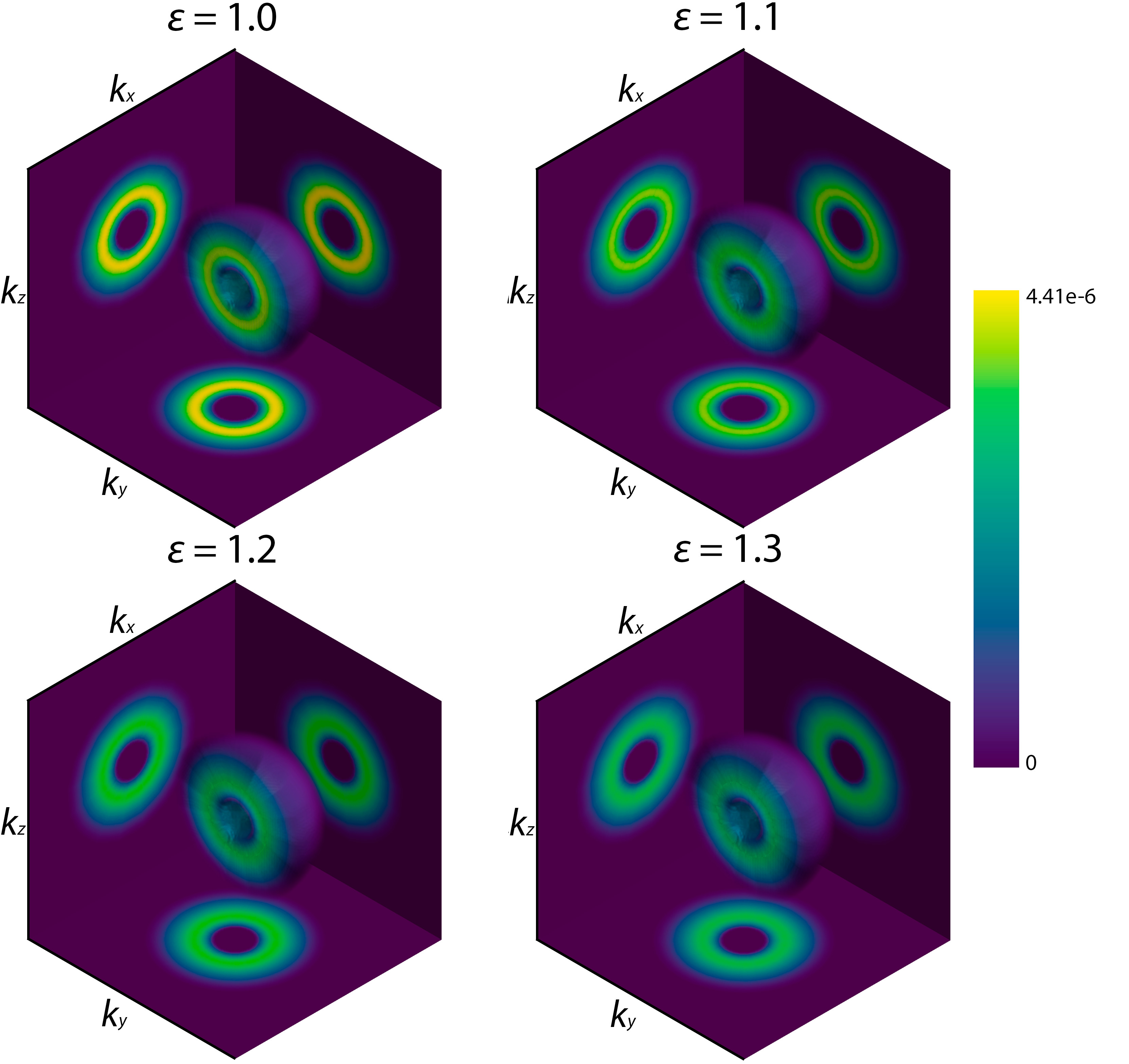}
    \caption{Pseudocolor plot of the angle-resolved differential probabilities of positron creation corresponding to the positron momentum components $k_z$, $k_x$, and $k_z$. The spectra are presented for collisions of the nuclei with $Z=96$ and $\varepsilon=1.0$, $1.1$, $1.2$, $1.3$ at $R_{\mathrm{min}}=17.5$~fm. Angular momentum projections $J_{z}=-1/2$, $1/2$ and $3/2$ are taken into account. For more details, please see the text.}
    \label{fig:3d_distributions_z_96}
\end{figure}

In Figs.~\ref{fig:3d_distributions_z_92} and \ref{fig:3d_distributions_z_96}, three-dimensional angle-resolved energy distributions of outgoing positrons are shown for collisions of two identical nuclei with the charge numbers $Z=92$ and $Z=96$, respectively, and scaled energy parameters $\varepsilon=1.0$, $1.1$, $1.2$, $1.3$. Angular momentum projections $J_{z}=-1/2$, $1/2$ and $3/2$ were included in the calculations. The angle-resolved differential positron creation probabilities were calculated according to Eq.~(\ref{eq:distr}) at the end of the time propagation in the rotating frame and then transformed to the initial center-of-mass frame of reference, thus 
the momentum components in Figs.~\ref{fig:3d_distributions_z_92} and \ref{fig:3d_distributions_z_96} refer to the momentum space associated with the initial direction of the coordinate axes. In the figures, only a half of the three-dimensional distribution is presented as a hemisphere to provide a more detailed view of the region where the distribution has its maximum. Two-dimensional slices of the three-dimensional distribution are shown on the $k_z-k_x$, $k_x-k_y$ and $k_z-k_y$ planes, each passing through the center of the three-dimensional distribution and parallel to the respective plane.

The angle-resolved energy distributions of outgoing positrons in Figs.~\ref{fig:3d_distributions_z_92} and \ref{fig:3d_distributions_z_96} do not manifest any significant anisotropy. In Ref.~\cite{Dulaev_2024_Angular}, two-center calculations without the rotational coupling term demonstrated that the two-dimensional energy--angle distributions obtained by integration over the azimuthal $\phi_k$ angle are nearly spherically symmetric. In the present study, we calculated three-dimensional distributions, which account for the rotational coupling. Comparing the present and previous results, we can conclude that the rotational coupling at the collision energies $6-8$~MeV/u currently under consideration does not break the isotropy of the angular distributions of emitted positrons. The brightness of the pseudocolor plots, which reflects the value of the differential probability, decreases with increasing $\varepsilon$ for both $Z=92$ and $Z=96$, indicating a decrease of the differential probability. This observation is consistent with the behavior seen in the supercritical collision regime for the  total probabilities and angle-integrated energy distributions.

\section{Conclusion}
In the present work, the positron production in slow collisions of heavy nuclei has been studied within the two-center approach with the rotational coupling taken into account. The time-depended Dirac equation for positrons has been solved with the help of the generalized pseudospectral method in modified prolate spheroidal coordinates. The rotational coupling term, which emerges in the time-dependent Dirac equation after transformation to the rotating frame of reference and causes transitions between the states with different angular momentum projections, is explicitly included in the calculations. The total pair-creation probabilities as well as the angle-integrated energy distributions of outgoing positrons are obtained. We have also calculated three-dimensional angle-resolved energy distributions of the positrons by projecting the wave packet of emitted positrons on the relativistic plane waves. The impact of the rotational coupling on the total positron creation probabilities, as well as angle-integrated and angle-resolved energy distributions is assessed. The results of our calculations show little influence of the rotational coupling for the collision energies under consideration ($6-8$~MeV/u). The rotational coupling does not alter the signatures of the supercritical regime in positron production for all the total probabilities, angle-integrated and angle-resolved energy distributions either.

\begin{acknowledgments}
The two-center calculations in prolate spheroidal coordinates were supported by the Russian Science Foundation (Grant No 22-62-00004, \cite{rscf}). The one-center calculations beyond the monopole approximations were supported by the Theoretical Physics and Mathematics Advancement Foundation ``BASIS'' (Grant No 23-1-1-54-3).
\end{acknowledgments}

\bibliography{main.bib}
\end{document}